\documentclass[prd,twocolumn,superscriptaddress,floatfix,nopacs,preprintnumbers,nofootinbib]{revtex4-2}
\usepackage[utf8]{inputenc}

\usepackage[caption = false]{subfig}
\usepackage[normalem]{ulem}
\usepackage{xcolor}

\usepackage{mathrsfs}
\usepackage{amssymb,bm}
\usepackage{amsmath}
\usepackage{mathtools}
\usepackage{slashed}
\usepackage{graphics}
\usepackage{graphicx}
\usepackage{physics}

\newcommand{\vect}[1]{\boldsymbol{#1}_{\perp}}
\newcommand{\nc}{N_\mathrm{c}}

\newcommand{\gev}{\mathrm{GeV}}

\newcommand{\bt}{\vect{b}}
\newcommand{\rt}{\vect{r}}

\newcommand{\pt}{\vect{p}}

\newcommand{\kt}{\vect{k}}
\newcommand{\xt}{\vect{x}}
\newcommand{\yt}{\vect{y}}

\newcommand{\lqcd}{\Lambda_\mathrm{QCD}}

\usepackage[breaklinks,colorlinks,citecolor=citcolor,urlcolor=blue,linkcolor=lcolor]{hyperref}
\usepackage[capitalise]{cleveref}

\definecolor{lcolor}{rgb}{0.5,0,0}
\definecolor{citcolor}{rgb}{0,0.3,0.0}

\newcommand{\as}{\alpha_\mathrm{s}}

\begin{document}

\author{Heikki M\"{a}ntysaari}
\email{heikki.mantysaari@jyu.fi}
\author{Yossathorn Tawabutr}
\email{yossathorn.j.tawabutr@jyu.fi}
\affiliation{
Department of Physics, University of Jyv\"askyl\"a,  P.O. Box 35, 40014 University of Jyv\"askyl\"a, Finland
}
\affiliation{
Helsinki Institute of Physics, P.O. Box 64, 00014 University of Helsinki, Finland
}
\title{Complete Next-to-Leading Order Calculation of Single Inclusive $\pi^0$ Production in Forward Proton-Nucleus Collisions}

\begin{abstract}
    We present the first fully consistent calculation of inclusive $\pi^0$ production at forward rapidities at next-to-leading order (NLO) accuracy in proton-lead collisions at $\sqrt{s}=8.16$ TeV within the Color Glass Condensate approach. The center-of-mass energy dependence is determined by  the Balitsky-Kovchegov equation with the initial condition constrained at NLO accuracy by the proton structure function measurements. We find that the LHCb data further constrain this non-perturbative input, and that both the NLO corrections to the evolution equation and to the impact factor  have significant effects on the nuclear modification factor $R_{p\mathrm{Pb}}$. The complete NLO calculation is shown to be in qualitative agreement with the LHCb data, and the predictions are found to be sensitive to the details of the applied deep inelastic scattering (DIS) fit. These results demonstrate the need to perform global analyses of DIS and LHC data in order to probe gluon saturation phenomena at precision level at very small momentum fraction $x$.

\end{abstract}

\maketitle 

\section{Introduction}

Proton-nucleus ($p$A) collisions at high center-of-mass energies studied at RHIC and especially at the LHC provide access to the part of nuclear wave function with very small momentum fraction $x$. This, together with the fact that densities in heavy nuclei are enhanced by a factor $\sim A^{1/3}$, makes forward particle production measurements in p+A collisions a promising channel to probe non-linear QCD dynamics and look for signatures of gluon saturation~\cite{Morreale:2021pnn}.

Inclusive particle production measurements at the LHC both at midrapidity~\cite{ALICE:2012mj,ALICE:2018vhm} and at forward rapidities~\cite{LHCb:2021vww,LHCb:2022tjh}, together with the corresponding measurements at RHIC in forward kinematics~\cite{STAR:2006dgg,BRAHMS:2004xry,PHENIX:2011puq}, have shown clear modifications in the p+A cross section compared to the scaled proton-proton results. These modifications take place at transverse momenta of a few GeV, which is of the same order as the expected saturation scale for heavy nuclei at LHC energies. Thus, an accurate description of nuclear effects seen at the LHC is a crucial test for the gluon saturation picture at the highest collider energies available. To describe QCD in the high-density domain where saturation effects are important, an effective theory of QCD known as the Color Glass Condensate (CGC)~\cite{Gelis:2010nm,Kovchegov:2012mbw} has been developed. It is the theoretical framework used in this paper, and leading order (LO) CGC calculations~\cite{Dumitru:2005gt,Albacete:2010bs,Tribedy:2010ab,Lappi:2013zma,Ducloue:2016ywt,Mantysaari:2019nnt,Ducloue:2015gfa,Ducloue:2016pqr,Albacete:2016tjq,Albacete:2012xq} have been successful in describing, at least qualitatively, different particle production spectra from  RHIC and LHC. 

Quantitative comparisons to the current and upcoming precision data require theoretical CGC calculations to be promoted to the NLO accuracy. In recent years, there has been a rapid progress in the field that has made NLO calculations in the saturation region possible. The crucial developments in the context of this work include the derivation of the NLO cross section for the single inclusive particle production obtained in Refs.~\cite{Chirilli:2012jd,Chirilli:2011km}, which are based on the leading-order formalism developed in~\cite{Dumitru:2005gt,Albacete:2010bs}. In the first phenomenological application, uncontrolled large NLO corrections were found rendering the cross section negative~\cite{Stasto:2013cha}. This issue triggered a lot of attention in the community~\cite{Ducloue:2016shw,Iancu:2016vyg,Ducloue:2017dit,Watanabe:2015tja,Xiao:2018zxf,Shi:2021hwx,Altinoluk:2014eka}. To date, two solutions have been proposed. The first is through threshold resummation \cite{Xiao:2018zxf}, which has led to an agreement with LHCb~\cite{LHCb:2021vww} and BRAHMS~\cite{BRAHMS:2004xry} data in Ref.~\cite{Shi:2021hwx} (however no NLO corrections to the high-energy evolution were included in~\cite{Shi:2021hwx}). The second solution to the negativity problem is by modifying the factorization procedure as shown in Ref.~\cite{Iancu:2016vyg}. This solution was extended to include the running coupling in~\cite{Ducloue:2017dit}.  

In CGC calculations, the $q\bar q$ dipole-target scattering amplitude, a.k.a. the dipole amplitude, is a convenient degree of freedom used to describe interactions with the target at high energies. The center-of-mass energy dependence of the dipole amplitude, and that of the cross section, can be calculated perturbatively by solving the Balitsky-Kovchegov (BK) equation~\cite{Kovchegov:1999yj,Balitsky:1995ub}, which is also available at NLO accuracy~\cite{Balitsky:2008zza,Lappi:2016fmu,Lappi:2015fma}. However, the starting point of the evolution equation requires a non-perturbative input describing the structure of the target proton or nucleus at moderate momentum fraction $x$. 
Typically parametrizations based on the McLerran-Venugopalan (MV) model~\cite{McLerran:1993ka,McLerran:1993ni,McLerran:1994vd} (which is strictly speaking valid for very large nuclei) have been used. 
The free parameters for the proton case have been extracted from fits to the HERA structure function data~\cite{H1:2009pze,H1:2015ubc}, and such fits at NLO accuracy became possible
when the CGC results for the total deep inelastic scattering (DIS) cross section at NLO were obtained in Refs.~\cite{Beuf:2017bpd,Hanninen:2017ddy,Beuf:2021srj}. Recently, an alternative approach to obtain the initial condition for the BK evolution has been proposed in Refs.~\cite{Dumitru:2020gla,Dumitru:2023sjd}, where this initial condition is obtained based on the  proton wave function in the valence region, including also the first perturbative correction from a  gluon emission.

The purpose of this work is to  perform the first fully consistent NLO calculation of inclusive $\pi^0$ production in LHC kinematics within the CGC framework. The initial condition for the small-$x$ evolution at NLO is obtained from NLO fits to HERA data performed in Ref.~\cite{Beuf:2020dxl}. The determined dipole-proton amplitude is generalized to the dipole-nucleus case by employing the optical Glauber model~\cite{Lappi:2013zma}, and as such there is no free parameter describing the nuclear high-energy structure despite the standard Woods-Saxon geometry. The $\pi^0$ production cross section is then calculated at the NLO accuracy by adopting the prescription for the factorization procedure and running coupling prescription developed in Refs.~\cite{Iancu:2016vyg,Ducloue:2017dit}.

This work is structured as follows. In Sec.~\ref{sect:setup}, we present the calculation of inclusive $\pi^0$ production cross section at next-to-leading order within the CGC framework. The cross section is expressed in terms of the dipole-target scattering amplitude, whose initial condition and energy evolution are discussed in Sec.~\ref{sect:dip}. The results at the parton and hadron levels, together with  comparisons to the LHCb data, are shown in Secs.~\ref{sec:partonlevel} and~\ref{sect:hadron_lvl_res}, respectively. Finally, we present our conclusions in Sec.~\ref{sect:conclusion}. A detailed discussion on the sensitivity of the results to the running coupling prescription in the hard impact factor is presented in Appendix.~\ref{sect:app_rc}.

\section{Particle production in proton-nucleus collisions}\label{sect:setup}

The forward particle production cross section in $p$A collisions at next-to-leading order is calculated in~\cite{Chirilli:2012jd,Chirilli:2011km}, where the cross section is shown to factorize to a hard part $\mathcal{H}_{a\to c}$ (a.k.a. impact factor), the part describing the eikonal interaction of partons with the target color field $S_{a,c}$, and the collinearly factorized parton distribution function $f_a$ and fragmentation function $D_{h,c}$:
\begin{align}\label{CXYfact}
    %&\frac{\dd\sigma^{p+A\to h+X}}{\dd y \dd[2]\pt} = \sum_a \int \frac{\dd[z]}{z^2} \frac{\dd{x}}{x} \xi x f_a(x,\mu^2) D_{h,c}(z,\mu) \\
    %&\times \int [\dd{\xt}] S_{a,c}([\xt]) \mathcal{H}_{a\to c} \\
    &\frac{\dd\sigma^{p+A\to h+X}}{\dd y \, \dd[2]\pt} = \sum_{a,c} f_a \otimes \mathcal{H}_{a\to c} \otimes S_{a,c} \otimes D_{h,c}\,.
\end{align}
The process described in Eq.~\eqref{CXYfact} involves a parton $a$ from the incoming proton interacting with the nuclear target $A$, resulting in parton $c$ that comes out and fragments into the observed hadron $h$. In this work, we include gluons and light quarks as we consider only $\pi^0$ production. The quark mass corrections are not included in the NLO impact factor~\cite{Chirilli:2012jd,Chirilli:2011km} nor in the NLO DIS fit~\cite{Beuf:2020dxl}. 

Since the interaction with the nuclear target is eikonal, the shockwave picture applies. This allows $S_{a,c}$ to be expressed in terms of Wilson lines describing the eikonal propagation of partons in the target color field. In the large-$\nc$ limit employed throughout this work~\cite{tHooft:1973alw}, each term of $S_{a,c}$, involving Wilson line correlators, can be written in terms of the dipole-target scattering matrix which is a correlator of two fundamental light-cone Wilson lines $V_{\xt}$:
\begin{align}\label{dip_amp}
    S(\rt,\bt, X) = \frac{1}{\nc} \left \langle \tr\left[ V_{\bt+\rt/2}V^\dagger_{\bt-\rt/2}\right] \right \rangle_X,
\end{align}
where
\begin{align}
    V_{\xt} = \mathcal{P}\exp\left[ig\int_{-\infty}^{\infty}\dd{x^-} A^+(0^+,x^-,\xt)\right] .
\end{align}
Here, $\mathcal{P}$ is a path ordering operator and $A^+$ is the classical color field, with the light-cone coordinates defined such that $v^{\pm}=(v^0\pm v^3)/\sqrt{2}$. The proton is moving predominantly in the light-cone minus direction, while the target nucleus is moving in the plus direction. In Eq.~\eqref{dip_amp}, the angle brackets, $\left\langle \cdots \right\rangle_X$, denote the average taken over the target color field configurations. The correlator is evolved in Bjorken $x$ from the initial $X_0$ down to $X$ by solving the BK equation.  

The hard impact factor, $\mathcal{H}_{a\to c}$, includes perturbative corrections that can be calculated order-by-order.
At the leading order, the incoming parton $a$ and the fragmenting parton $c$ are the same, $a=c$. This gives two possible channels of interaction: ``quark channel'' and ``gluon channel'', corresponding to whether $a$ is a quark or a gluon. At NLO the quark can emit a gluon, or the gluon can split to a $q\bar q$ or $gg$ pair. This leads to four possible channels at NLO: $qq$ channel, $qg$ channel, $gq$ channel and $gg$ channel, corresponding to $a$ and $c$, respectively. The ``primary parton'' denoted by $c$ that carries the momentum fraction $\xi$ relative to $a$ fragments into the hadron $h$. The phase space of the other unobserved parton is integrated over. The transverse integrals result in collinear divergences that are subtracted through the DGLAP evolution to the parton distribution functions (PDFs) and fragmentation functions (FFs). For consistency, we also use the PDFs and FFs at NLO. Specifically, throughout this work, we employ the MSTW PDFs~\cite{Martin:2009iq} and the DSS fragmentation functions~\cite{deFlorian:2007aj}. 

In the hard factor, there is also an integral over the longitudinal momentum fraction, $\xi$, of the primary parton. This integral contains rapidity divergent terms that can be included in the BK evolution of the target. In this work we apply the ``unsubtracted scheme'' as in Ref.~\cite{Ducloue:2017dit} (and e.g. in related NLO calculations in Refs.~\cite{Beuf:2020dxl,Mantysaari:2022bsp,Mantysaari:2022kdm,Hanninen:2022gje}), in which the rapidity divergence is not explicitly subtracted from the NLO contribution of $\mathcal{H}_{a\to c}$. Specifically, dipole amplitudes in the LO term are evaluated at the initial scale, $X_0$, of the small-$x$ evolution. At NLO, dipoles are evaluated at
\begin{equation}\label{Xxi}
    X(\xi) = \frac{X_g}{1-\xi},
\end{equation}
where $0<\xi<1-X_g$ is the fraction of the primary parton longitudinal momentum carried by the fragmenting parton, and the fraction of the target longitudinal momentum transferred in the process is 
\begin{equation}
\label{eq:xg}
    X_g = \frac{k_{\perp}}{\sqrt{s}}\,e^{-y}\,.
\end{equation}
Here $\kt$ is the transverse momentum of the produced quark or gluon, and $k_{\perp} = |\kt|$. The kinematical upper bound $\xi<1-X_g$ ensures that the integral over $\xi$ in the NLO impact factor does not diverge.
%%%%%%%%%%%%%%%%%%%%%%%%%%%
This choice of limit is in accordance with the DIS fits~\cite{Beuf:2020dxl} whose results are employed in the calculation of the dipole amplitudes in this work (see discussion in Sec.~\ref{sect:dip}). As our calculation is in the LHC kinematics where $X_g\sim 10^{-5}$, the sensitivity to the large-$x$ region is small.
%%%%%%%%%%%%%%%%%%%%%%%%%%%%
Before we proceed to the actual calculation, we introduce another important kinematic variable:  the fraction of proton longitudinal momentum carried by the incoming parton before the emission of the primary parton. The ratio is given by
\begin{align}
    &x_p = \frac{k_{\perp}}{\sqrt{s}}\,e^{y}\,.
\end{align}

In this scheme, at LO in the impact factor, the quark- and gluon-channel cross sections are given by \cite{Dumitru:2002qt,Dumitru:2005gt,Albacete:2010bs,Ducloue:2017dit}
\begin{widetext}
\begin{subequations}\label{LOexpr}
\begin{align}
    &\frac{\dd\sigma^{p+A\to h+X}}{\dd y \, \dd[2]\pt}\Big|_{\text{LO, q}} = \frac{1}{4\pi^2}\int_{\tau}^1\frac{\dd z}{z^2}\,D_{h/q}(z)\,x_p\,q(x_p) \int \dd[2]{\bt} \int \dd[2]{\rt} e^{-i\kt\cdot\rt}S(\rt,\bt,X_0) , \label{LOq} \\ 
%%%%%%%%%%%%%%%%%%%%%%%%%%%%%%%%%%%%%%%%%%%%%%%%%%%%%%%%%%%%%%%%%%%%%%%%%%%
    &\frac{\dd\sigma^{p+A\to h+X}}{\dd y \, \dd[2]\pt}\Big|_{\text{LO, g}} = \frac{1}{4\pi^2}\int_{\tau}^1\frac{\dd z}{z^2}\,D_{h/g}(z)\,x_p\,g(x_p) \int \dd[2]{\bt} \int \dd[2]{\rt} e^{-i\kt\cdot\rt} \left[S(\rt,\bt,X_0) \right]^2 ,  \label{LOg}  
\end{align}
\end{subequations}
\end{widetext}
respectively. Here, the lower limit, $\tau = (p_{\perp}/\sqrt{s})e^{y}$, is imposed to the fragmentation function integral so that $x_p\leq 1$, and  $p_{\perp} = zk_{\perp}$. Note that Eqs.~\eqref{LOexpr} are written in the large-$N_c$ limit. Furthermore, $\bt=(\xt+\yt)/2$ is the impact parameter and $\rt=\xt-\yt$. With proton targets (when calculating the cross section in proton-proton collisions), we assume, as in DIS fits~\cite{Beuf:2020dxl}, that the impact parameter dependence factorizes and replace 
\begin{align}
    \int\dd[2]{\bt}\to\frac{\sigma_0}{2}.
\end{align}
Here, the proton transverse area $\sigma_0/2$ is extracted in fits to proton structure function data. In the complete calculation that will sum over all channels, the light quark flavors in the quark-channel cross section \eqref{LOq} will be summed over, accounting for the fact that a quark (or an antiquark) of any flavor inside the proton can interact with the heavy nuclear target.

As for the NLO in the impact factor, the cross section for the $qq$ channel in the unsubtracted scheme has been calculated in Ref.~\cite{Ducloue:2017dit} based on Ref.~\cite{Chirilli:2012jd}. The explicit expression for the cross section reads
\begin{widetext}
\begin{align}\label{NLOqq}
    &\frac{\dd\sigma^{p+A\to h+X}}{\dd y \, \dd[2]\pt}\Big|_{\text{NLO, qq}} = \frac{N_c}{8\pi^2} \int_{\tau}^1\frac{\dd z}{z^2}\,D_{h/q}(z)  \left\{ \int_{\tau/z}^{1-X_g} \dd \xi \, \frac{x_p}{\xi}\,q\left(\frac{x_p}{\xi}\right)  \frac{1+\xi^2}{1-\xi} \int \dd[2]{\bt} \int \dd[2]{\rt} e^{-i\kt\cdot\rt} \right.   \\
    &\;\;\;\;\;\;\;\;\times \left[I_1(\rt, \bt, \xi) + I_1(\xi\rt, \bt, \xi) - 4I_2(\rt, \bt, \xi) + 2J(\rt, \bt, \xi)\right] \notag   \\
    &\;\;\;\;- \left. x_p\,q(x_p)\int_0^{1-X_g}\dd \xi \, \frac{1+\xi^2}{1-\xi} \int \dd[2]\bt \int \dd[2]{\rt} e^{-i\kt\cdot\rt} \left[2J_v(\rt, \bt,\xi) + \frac{\alpha_s}{\pi}\,S(\rt,\bt,X(\xi))\, \ln\frac{k^2_{\perp}(1-\xi)^2}{\mu^2}\right]  \right\}  , \notag
\end{align}
where
\begin{subequations}\label{NLOqqfuncs}
    \begin{align}
        J(\rt, \bt,\xi) &=  \int \dd[2]{\xt} \frac{\xt\cdot(\rt+\xt)}{x^2_{\perp}|\rt+\xt|^2}  \label{Jqq} \\
        &\;\;\;\;\;\times \frac{\alpha_s}{2\pi^2}\left[S(\rt+(1-\xi)\xt, \bt, X(\xi)) - S(\xi \xt, \bt, X(\xi))\,S(\rt+\xt, \bt, X(\xi))\right] , \notag \\ 
%%%%%%%%%%%%%%%%%%%%%%%%%%%%%%%%%%%%%%%%%%%%%%%%%%%%%%%%%%%%%%%%%%%%%%%%%%%
J_v(\rt, \bt,\xi) &=  \int \dd[2]\xt\frac{1}{x_{\perp}^2} \, \frac{\alpha_s}{2\pi^2}  \left[S(\rt-(1-\xi)\xt, \bt,X(\xi)) - S(\xt, \bt,X(\xi))\,S(\rt+\xi \xt, \bt,X(\xi))\right]    , \label{Jvqq} \\
%%%%%%%%%%%%%%%%%%%%%%%%%%%%%%%%%%%%%%%%%%%%%%%%%%%%%%%%%%%%%%%%%%%%%%%%%%%
I_1(\rt, \bt,\xi) &=  \frac{\alpha_s}{2\pi}\,S(\rt, \bt,X(\xi)) \,\ln\left[\frac{c_0^2}{r_{\perp}^2\mu^2}\right]   , \label{I1qq} \\
%%%%%%%%%%%%%%%%%%%%%%%%%%%%%%%%%%%%%%%%%%%%%%%%%%%%%%%%%%%%%%%%%%%%%%%%%%%
I_2(\rt, \bt,\xi) &=  \int \dd[2]{\xt} \frac{\xt\cdot(\rt+\xt)}{x_{\perp}^2|\rt+\xt|^2} \, \frac{\alpha_s}{4\pi^2} \, S((1-\xi)\xt-\xi \rt, \bt,X(\xi)) \,    .   \label{I2qq}  
    \end{align}
\end{subequations}
\end{widetext}
We perform a similar calculation to obtain the cross section for the other three remaining channels ($qg,gq,gg$) in the unsubtracted scheme starting from the results obtained in Ref.~\cite{Chirilli:2012jd}. The resulting expressions are 
\begin{widetext}
\begin{subequations}\label{NLOexpr}
\begin{align}
    &\frac{\dd\sigma^{p+A\to h+X}}{\dd y \, \dd[2]\pt}\Big|_{\text{NLO, gg}} = \frac{N_c}{4\pi^2} \int_{\tau}^1\frac{\dd z}{z^2}\,D_{h/g}(z)   \label{NLOgg} \\
    &\times  \left\{ \int_{\tau/z}^{1-X_g}\dd \xi \,\frac{x_p}{\xi}\,g\left(\frac{x_p}{\xi}\right) \frac{\left[1-\xi(1-\xi)\right]^2}{\xi(1-\xi)} \int \dd[2]{\bt} \int \dd[2]{\rt}  e^{-i\kt\cdot\rt}     \left[H_1(\rt,\bt,\xi) + H_1(\xi\rt,\bt,\xi) - H_2(\rt,\bt,\xi)\right] \right. \notag \\
&\;\;\;\;+ x_p\,g(x_p) \int_0^{1-X_g}\dd \xi \left[ \frac{2\xi}{1-\xi} + \xi(1-\xi) \right]  \int \dd[2]{\bt} \int \dd[2]{\rt}  e^{-i\kt\cdot\rt}  \left[H_3(\rt,\bt,\xi) - \frac{\alpha_s}{\pi} \left[S(\rt,\bt,X(\xi))\right]^2 \ln\frac{\kt^2(1-\xi)^2}{\mu^2}  \right]   \notag  \\
&\;\;\;\;+ \left. x_p\,g(x_p) \int_0^{1-X_g}\dd \xi   \left[\xi^2+(1-\xi)^2\right] \int \dd[2]\rt  e^{-i\kt\cdot\rt} \left[H_4(\rt,\bt,\xi)  - \frac{\alpha_s}{2\pi} \left[S(\rt,\bt,X(\xi))\right]^2 \ln\frac{\kt^2(1-\xi)^2}{\mu^2}  \right] \right\} , \notag  \\
%%%%%%%%%%%%%%%%%%%%%%%%%%%%%%%%%%%%%%%%%%%%%%%%%%%%%%%%%%%%%%%%%%%%%%%%%%%    
    &\frac{\dd\sigma^{p+A\to h+X}}{\dd y \, \dd[2]\pt}\Big|_{\text{NLO, qg}} = \frac{N_c}{4\pi^2} \int_{\tau}^1\frac{\dd z}{z^2}\,D_{h/g}(z) \label{NLOqg} \\
    &\times \int_{\tau/z}^{1-X_g} \dd \xi \, \frac{x_p}{\xi}\,q\left(\frac{x_p}{\xi}\right) \frac{1}{\xi}\left[1+(1-\xi)^2\right] \int \dd[2]{\bt} \int \dd[2]{\rt} e^{-i\kt\cdot\rt}  \left[ \frac{1}{2}I_1(\xi\rt,\bt,\xi) + \frac{1}{4}H_1(\rt,\bt,\xi) - K_1(\rt,\bt,\xi)\right]   , \notag \\
%%%%%%%%%%%%%%%%%%%%%%%%%%%%%%%%%%%%%%%%%%%%%%%%%%%%%%%%%%%%%%%%%%%%%%%%%%%
    &\frac{\dd\sigma^{p+A\to h+X}}{\dd y \, \dd[2]\pt}\Big|_{\text{NLO, gq}} = \frac{1}{4\pi^2}\int_{\tau}^1\frac{\dd z}{z^2}\,D_{h/q}(z) \label{NLOgq} \\
    &\times \int_{\tau/z}^{1-X_g}\dd \xi \, \frac{x_p}{\xi}\,g\left(\frac{x_p}{\xi}\right) \left[(1-\xi)^2 + \xi^2\right]   \int \dd[2]{\bt} \int \dd[2]\rt e^{- i\kt\cdot\rt}  \left[ \frac{1}{2}I_1(\rt,\bt,\xi) + \frac{1}{4}H_1(\xi\rt,\bt,\xi) - K_2(\rt,\bt,\xi) \right] , \notag 
\end{align}
\end{subequations}
where
\begin{subequations}\label{NLOfuncs}
    \begin{align}
        &H_1(\rt,\bt,\xi) = \frac{\alpha_s}{\pi} \left[S(\rt,\bt,X(\xi))\right]^2  \ln\frac{c_0^2}{\mu^2r_{\perp}^2}  \, ,     \label{H1gg} \\
%%%%%%%%%%%%%%%%%%%%%%%%%%%%%%%%%%%%%%%%%%%%%%%%%%%%%%%%%%%%%%%%%%%%%%%%%%%
&H_2(\rt,\bt,\xi) =  \int \dd[2]\xt\frac{\xt\cdot(\xt+\rt)}{x_{\perp}^2|\xt+\rt|^2}  \label{H2gg} \\
&\times \frac{2\alpha_s}{\pi^2} \, S(\xt,\bt,X(\xi)) \, S(\xi(\xt+\rt),\bt,X(\xi))\,S((1-\xi)\xt-\xi \rt,\bt,X(\xi)) \, , \notag \\
%%%%%%%%%%%%%%%%%%%%%%%%%%%%%%%%%%%%%%%%%%%%%%%%%%%%%%%%%%%%%%%%%%%%%%%%%%%
&H_3(\rt,\bt,\xi) =  \int \dd[2]\xt\frac{1}{x_{\perp}^2}  \label{H3gg} \\
&\times \frac{\alpha_s}{\pi^2}\left[ S(\xt,\bt,X(\xi))\,S(\rt-(1-\xi)\xt,\bt,X(\xi))\,S(\rt+\xi\xt,\bt,X(\xi)) - \left[S(\rt+(1-\xi)\xt,\bt,X(\xi))\right]^2  \right] , \notag \\
%%%%%%%%%%%%%%%%%%%%%%%%%%%%%%%%%%%%%%%%%%%%%%%%%%%%%%%%%%%%%%%%%%%%%%%%%%%
&H_4(\rt,\bt,\xi) =  \int \dd[2]\xt\frac{1}{x_{\perp}^2}   \label{H4gg} \\ 
&\times \frac{\alpha_s}{2\pi^2} \left[ S(\rt-\xi\xt,\bt,X(\xi))\,S(\rt+(1-\xi)\xt,\bt,X(\xi)) - \left[S(\rt+(1-\xi)\xt,\bt,X(\xi))\right]^2 \right]  , \notag \\
%%%%%%%%%%%%%%%%%%%%%%%%%%%%%%%%%%%%%%%%%%%%%%%%%%%%%%%%%%%%%%%%%%%%%%%%%%%
&K_1(\rt,\bt,\xi) = \int \dd[2]\xt \frac{\xt\cdot(\xt+\rt)}{x_{\perp}^2|\xt+\rt|^2} \, \frac{\alpha_s}{2\pi^2} \, S(\xt,\bt,X(\xi))\,S(\xi\rt-(1-\xi)\xt,\bt,X(\xi)) \, , \label{K1qg} \\ 
%%%%%%%%%%%%%%%%%%%%%%%%%%%%%%%%%%%%%%%%%%%%%%%%%%%%%%%%%%%%%%%%%%%%%%%%%%%
&K_2(\rt,\bt,\xi) =   \int \dd[2]\xt \frac{\xt\cdot(\xt+\rt)}{x_{\perp}^2|\xt+\rt|^2} \, \frac{\alpha_s}{2\pi^2} \, S(\xi\xt,\bt,X(\xi))\,S(\rt+(1-\xi)\xt,\bt,X(\xi)) \, . \label{K2gq} 
    \end{align}
\end{subequations}
\end{widetext}
Similarly to the LO case, in the actual cross-section calculation, every channel except for the $gg$ channel involves a sum over the light quark flavors.

In Eqs. \eqref{NLOqq} and \eqref{NLOexpr}, the strong coupling $\as$ is taken to run with the transverse momentum of the interacting parton throughout this work. This scheme choice was shown in Ref.~\cite{Ducloue:2017dit} to avoid the ``fake potential'' issue that would result in uncontrolled NLO correction. In Appendix~\ref{sect:app_rc}, we demonstrate that there is only a moderate sensitivity on the running coupling scheme that mostly cancels in cross section ratios, as long as we use one of the schemes that avoid the fake potential issue.

The $z$-integrals in Eqs.~\eqref{LOexpr}, \eqref{NLOqq}, and \eqref{NLOexpr} can be interpreted as the convolutions between the fragmentation function and the parton-level cross section, which is the single inclusive parton production cross section in $pA$ collisions. In some cases of the calculation below, the parton-level cross sections will be negative at given parton transverse momentum, $\kt$. This negativity problem results from the fact that already at leading order the cross section is proportional to the Fourier transform of the dipole amplitude (see Eq.~\eqref{LOexpr}), and this Fourier transform is not guaranteed to be positive definite.\footnote{Note that the negativity problem discussed here is separate from the negativity problem resolved in Refs.~\cite{Iancu:2016vyg,Ducloue:2017dit}, which can be encountered in the NLO calculation if the running coupling prescription and the factorization between the NLO impact factor and the high-energy evolution are not properly handled.} We will return to this discussion in Sec.~\ref{sect:dip}. To avoid propagating the negative cross section problem to the hadron level, we replace the parton-level cross section, $\dd\sigma_{\text{parton}}$, by $\max\{0,\dd\sigma_{\text{parton}}\}$ in the integrand of all the fragmentation-function integrals over $z$ that appear in Eqs.~\eqref{LOexpr}, \eqref{NLOqq}, and \eqref{NLOexpr}. We have confirmed numerically that this has a negligible effect on the hadron-level cross section except at $p_{\perp}$ values where the hadron-level results would be negative.

\section{Dipole amplitude and the high-energy evolution}\label{sect:dip}

As we see from Eqs.~\eqref{NLOqq} to \eqref{NLOfuncs}, each term in the NLO cross section involves dipole amplitudes, $S(\rt,X(\xi))$, for some transverse separation $\rt$. In the shockwave picture, these amplitudes correspond to the interactions between the color-singlet quark-antiquark dipole and the target nucleus. The dipoles are evolved via the small-$x$ BK evolution equation from the initial value, $X_0$, to the smaller value of $X(\xi)$ given in Eq.~\eqref{Xxi}.   

In the  case where the nuclear target is in fact a proton, we adopt the parametrization employed in \cite{Beuf:2020dxl} (and previously in leading order fits~\cite{Albacete:2010sy,Lappi:2013zma}), where we take the non-perturbative initial conditions for the dipole-proton scattering amplitude to be
\begin{align}\label{ICdip_pp}
    S(\rt, x=X_{0}) &= \exp\left[ -\frac{(r_{\perp}^2Q_{s,0}^2)^\gamma}{4}\ln \left( \frac{1}{r_{\perp} \lqcd} + e\right)\right].  
\end{align}
Here $\lqcd$ is taken to be 0.241 GeV, while $\gamma$ and $Q_{s,0}^2$ are free parameters that are determined by performing a fit to the proton structure function data from HERA~\cite{H1:2009pze,H1:2015ubc}. The determined dipole-proton scattering amplitudes are available at~\cite{Beuf:2020dxl,heikki_mantysaari_2020_4229269}. Physically, $Q_{s,0}^2$ controls the saturation scale at the initial condition and $\gamma$ is the anomalous dimension which determines the shape of the dipole amplitude in the limit of small dipole size $r_{\perp}$. In addition, other free parameters include the proton transverse area, $\sigma_0/2$, and a parameter controlling the scale of the running coupling in coordinate space. The latter is denoted by $C^2$ in Refs.~\cite{Albacete:2010sy,Lappi:2013zma,Beuf:2020dxl}, see Eq.~\eqref{aspos} for an explicit expression for the $\as$ as a function of transverse distance scale.

The first fit to determine the required non-perturbative input at next-to-leading order accuracy was performed in Ref.~\cite{Beuf:2020dxl}, and the determined dipole amplitudes are also used in this work. In principle, in the fully consistent NLO analysis, one should also use the NLO BK evolution equation~\cite{Balitsky:2008zza}. As the NLO BK equation is computationally challenging, it was approximated in Ref.~\cite{Beuf:2020dxl} by the leading order BK equation together with the running coupling corrections and a resummation of the most important higher order corrections enhanced by large transverse logarithms. This imposes a dependence on the resummation scheme that specifies how the large transverse logarithms that appear in the collinear regime of the NLO corrections are resummed. It is worth noting that this collinear resummation scheme employed to approximate the NLO BK equation is distinct and separate from the ``subtraction scheme'' of the cross section, which is discussed in Sec.~\ref{sect:setup} and Ref.~\cite{Ducloue:2017dit}. The three different resummation schemes, which lead to three different approximations of the NLO BK evolution considered here, are: the kinematically constraint BK (KCBK) equation~\cite{Beuf:2014uia}, the rapidity local resummed BK (ResumBK) equation~\cite{Iancu:2015joa,Iancu:2015vea}, and the target-momentum BK (TBK) equation \cite{Ducloue:2019ezk}. The TBK equation is formulated in term of the target rapidity, $\eta$, which is then transformed to the projectile rapidity used to derive the impact factor as in Ref.~\cite{Beuf:2020dxl}.  Here, the terminology is the same as in Ref.~\cite{Beuf:2020dxl}, and we note that in Ref.~\cite{Hanninen:2021byo} it was found that the remaining NLO corrections not enhanced by these large transverse logarithms have very small effects on DIS cross sections. These evolution equations have been successfully employed in NLO calculations of exclusive vector meson~\cite{Mantysaari:2022bsp,Mantysaari:2022kdm} and heavy quark~\cite{Hanninen:2022gje} production in DIS.

In Ref.~\cite{Beuf:2020dxl}, many different fits corresponding to different starting points of the BK evolution and different running coupling schemes are used. Although the running coupling prescriptions employed in~\cite{Beuf:2020dxl} do not exactly correspond to the ones used in the impact factor, the effect of the discrepancy is formally of higher orders in $\as$. Additionally, the reduced cross section data were fitted separately from the pseudodata consisting of an estimated light quark contribution. In this work, we only consider fits to the reduced cross section data, as all fits to the light quark pseudodata have a large anomalous dimension, $\gamma>1$. This could make the cross section negative, as discussed later in this Section. Of the fits to the reduced cross section data, we use the ones for which the evolution starts at $X_0=0.01$ (above which the dipole is frozen at its initial condition), unless otherwise specified.   

For the dipole-nucleus scattering process, it is not possible to perform a similar fit to the structure function data to obtain the scattering amplitude due to the lack of small-$x$ nuclear-DIS data. Instead, we generalize the dipole-proton amplitude to the dipole-nucleus case using the optical Glauber model as in Ref.~\cite{Lappi:2013zma}. This corresponds to modifying the initial saturation scale from Eq.~\eqref{ICdip_pp} to account for the impact-parameter profile of the heavy nucleus. The dipole-nucleus amplitude that we use as an initial condition for the BK evolution then reads
\begin{align}\label{ICdip_pA}
    S_{A}(\rt,\bt, X_{0}) &= \exp\left[- \frac{\sigma_0}{2} \, AT_A(\bt) \, \frac{(r_{\perp}^2Q_{s,0}^2)^\gamma}{4} \right. \\ 
    &\;\;\;\;\;\;\;\;\times \left.\ln \left( \frac{1}{r_{\perp}\lqcd} + e\right)\right]. \notag
\end{align}
The free parameters ($\sigma_0/2, Q_{s,0}^2$ and $\gamma$) are exactly the same as in the dipole-proton amplitude, and thus there is no freedom when moving to nuclear targets. Note that in the dilute regime the dipole-nucleus cross section is exactly $A$ times the dipole-proton one, and $S\to 0$ in the saturation regime. Here, $A$ is the mass number of the target, and $T_A(\bt)$ is the transverse thickness function of the nucleus, which can be obtained from the Woods-Saxon distribution of nuclear density. It is normalized such that $\int \dd[2]\bt T_A(\bt)=1$. The small-$x$ evolution is then applied to the initial condition \eqref{ICdip_pA} separately for each $b_{\perp}$, resulting in different dipole-nucleus interaction amplitudes as functions of $x$ and $r_\perp$ for different values of $b_{\perp}$. This method generalizing the dipole-proton amplitudes into their dipole-nucleus counterparts has been employed in \cite{Lappi:2021oag} for vector meson production at NLO in $\gamma A$ collisions. Furthermore, a successful description of  forward particle production cross section in proton-nucleus collisions have been obtained in leading order calculations in Refs.~\cite{Lappi:2013zma,Ducloue:2016pqr,Ducloue:2015gfa,Mantysaari:2019nnt}.

Another important characteristic of the fit in \cite{Beuf:2020dxl} is the fact that the anomalous dimension, $\gamma$, is allowed to exceed one. This is a similar practice to that employed in the preceding LO fits~\cite{Lappi:2013zma,Albacete:2010sy}. Unlike the total DIS cross section that only depends on the dipole amplitude convoluted with the photon wave function, the inclusive parton production cross section is proportional to the two-dimensional Fourier transform of the dipole amplitude. This Fourier transform is highly sensitive to the detailed shape of the dipole amplitude at very small transverse dipole size, $r_{\perp}$, which is the region that contributes minimally to the total DIS cross section. In particular, with $\gamma>1$, the Fourier transform is not positive definite~\cite{Giraud:2016lgg}, and as a result the parton level cross section can be negative. In order to minimize this issue, we mostly focus on fits reported in Ref.~\cite{Beuf:2020dxl} for which the anomalous dimension is $\gamma \approx 1$. An additional advantage of fits that have $\gamma\approx 1$ is the fact that only in this case does the nuclear saturation scale obey the natural scaling, $Q_{s,0,A}^2=\frac{\sigma_0}{2}A T_A(\bt) Q_{s,0}^2$. Furthermore, as mentioned at the end of Section \ref{sect:setup}, in the regions of parton transverse momentum, $k_{\perp}$, where the parton-level cross section is still negative, we replace these negative cross sections by zero when evaluating the convolution with fragmentation functions.

\section{Nuclear modification at parton level}
\label{sec:partonlevel}

In order to determine how the NLO corrections in the evolution equation and in the impact factor affect  the single inclusive cross section, we first calculate inclusive parton production cross section in the LHC kinematics. This means that the fragmentation function integral is not included. At this level, the $gg$ channel dominates in the spectra, followed by the $qq$ channel. While we will include all channels when calculating the hadron-level results in Sec.~\ref{sect:hadron_lvl_res}, the parton level results are shown only for these two channels to illustrate the importance of the various NLO corrections. In this Section, we also focus on the most central collisions only, i.e. $b_\perp=0$ in the applied optical Glauber model, as in this case the nuclear saturation scale is the highest and nuclear effects are most pronounced. However, we note that the $b_\perp=0$ collisions do not correspond to the experimental definition of most central $pA$ collisions. In Sec.~\ref{sect:hadron_lvl_res}, we will present results at hadron level for minimum bias collisions in which case the applied optical Glauber model (integrated over all $\bt$) is expected to be applicable. As the genuine parton-level cross section is not finite, we calculate (as in Ref.~\cite{Ducloue:2017dit}) only the part that does not involve the $1/\varepsilon$ divergence cancelled by the renormalization of the fragmentation function.

This exercise is motivated by the fact that the (approximative) next-to-leading order BK evolution employed in this work results in a different shape of the dipole-target scattering amplitude compared to the leading order evolution. More precisely, the leading order BK evolution drives the solution towards an asymptotic shape $N=1-S\sim r_\perp^{2\gamma}$ with $\gamma\sim 0.6\dots 0.8$ at small dipoles. This evolution removes the Cronin enhancement at moderate transverse momenta that appears in leading order calculations when an MV model initial condition is used~\cite{Albacete:2003iq}. On the other hand, the NLO evolution does not significantly modify the anomalous dimension, $\gamma$, i.e. the shape of the dipole amplitude in the very small $r_\perp$ region remains relatively close to that of the initial condition~\cite{Lappi:2016fmu,Iancu:2015joa,Iancu:2015vea,Ducloue:2019ezk}.

\begin{figure}
    \centering
    \includegraphics[width = \columnwidth]{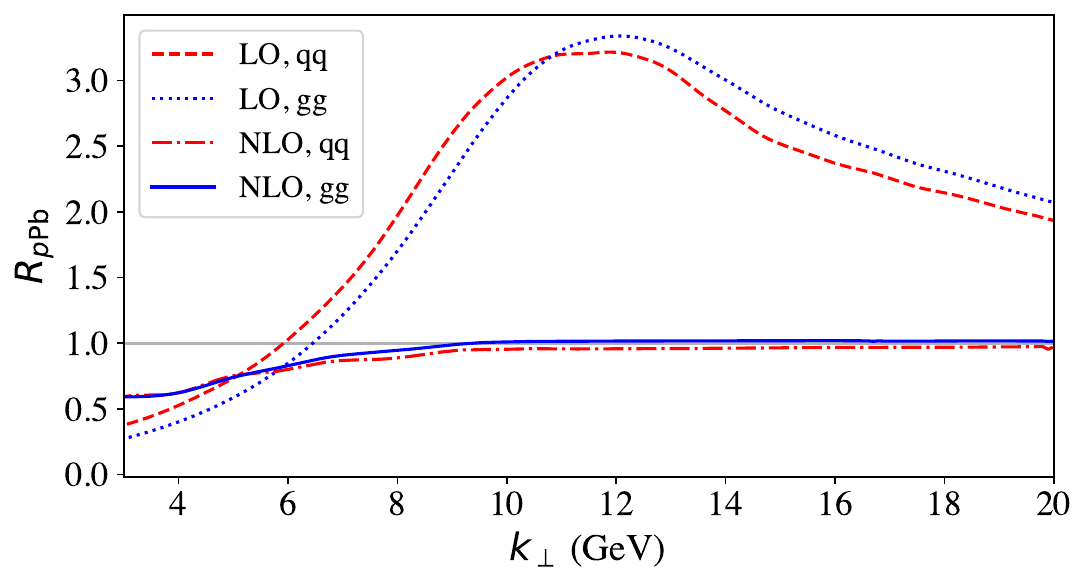}
    \caption{Nuclear modification factor in the dominant $qq$ and $gg$ channels for the most central proton-lead collisions calculated at the LHCb kinematics, $\sqrt{s}=8.16$ TeV and $y = 3$, as a function of the gluon transverse momentum.
    The KCBK resummation scheme is employed to approximate the complete NLO BK evolution, together with the parent-dipole running coupling prescription. The labels, LO and NLO, refer to the order in the impact factor.}
    \label{fig:parton_rpa}
\end{figure}

The nuclear modification factors shown in Fig.~\ref{fig:parton_rpa} for the quark and gluon production at most central collisions ($b_{\perp}=0$) are defined as 
\begin{equation}
    \label{RpA_parton}
    R_{p\mathrm{Pb}}^{qq,gg} = \frac{\dd N^{pA}_{qq,gg}}{N_\mathrm{bin}(\bt)\, \dd N^{pp}_{qq,gg}}\,.
\end{equation} 
Here,
\begin{align}
    &N_\mathrm{bin}(\bt)=A T_A(\bt) \sigma_\mathrm{inel} 
\end{align}
is the number of binary nucleon-nucleon collisions computed from the optical Glauber model with $\sigma_\mathrm{inel}$ referring to the total inelastic nucleon-nucleon cross section. The invariant yield in proton-nucleus collisions, $\dd N^{pA}_{qq,gg}$, is obtained by not performing the $\dd[2]\bt$ integral in expressions \eqref{NLOqq} and \eqref{NLOexpr} shown in Sec.~\ref{sect:setup} and evaluating the dipole amplitudes at fixed $b_\perp=0$. On the other hand, the proton-proton yield is obtained from the proton-proton cross section as~\cite{Lappi:2013zma}
\begin{equation}
    \dd N^{pp}_{qq,gg} = \frac{1}{\sigma_\mathrm{inel}} \, \dd \sigma^{pp}_{qq,gg} \, .
    \end{equation}
With all these ingredients, the quantity in Eq.~\eqref{RpA_parton} compares the $pA$ yield to its $pp$ counterpart and becomes $1$ when the $pA$ collisions can be seen as an incoherent superposition of independent proton-proton collisions. Note that the dependence on $\sigma_\mathrm{inel}$ cancels in Eq.~\eqref{RpA_parton}. To demonstrate the effect of NLO impact factor, we compare the full NLO calculation to the case where the LO impact factor is used with the NLO dipole. The latter case corresponds to Eq.~\eqref{LOexpr} with the dipole amplitude evolved to $X_g$. In both cases, the factorization scale for the PDFs is set to $\mu=4k_{\perp}$, where $\kt$ is again the transverse momentum of the produced parton. We use the KCBK resummation scheme for the BK evolution with the parent dipole running coupling prescription.

Since the NLO evolution does not evolve the dipole towards an asymptotic shape with a small anomalous dimension $\gamma$, we obtain a large Cronin enhancement peaked at $k_{\perp} \approx 12\,\gev$ in the forward LHC kinematics when the hard impact factor is calculated at the leading order. As discussed above, with a leading-order evolution one does not obtain a Cronin peak in the LHC kinematics~\cite{Lappi:2013zma,Albacete:2003iq}. Hence, we conclude that the NLO corrections to the small-$x$ evolution (not included in previous NLO calculations for forward particle production~\cite{Ducloue:2017dit,Shi:2021hwx})  can have a qualitatively significant effect on the nuclear modification factor. This Cronin peak vanishes when the NLO impact factor is used together with the NLO evolved dipole, highlighting the importance of performing the calculation consistently at the NLO accuracy. We note that no Cronin peak is seen in the forward LHCb~\cite{LHCb:2022tjh,LHCb:2021vww} or midrapidity ALICE~\cite{ALICE:2012mj,ALICE:2018vhm} data.

Furthermore, we find a strong nuclear suppression at low parton transverse momenta. The NLO results have a significantly weaker dependence on the gluon $k_{\perp}$ compared to the case where the leading order impact factor is used. This is because the latter corresponds to the $1\to 1$ kinematics, in contrast to the $1\to 2$ kinematics resulting from a hard  parton emission vertex at NLO. In the NLO case, the unobserved parton carries a significant transverse momentum, rendering the $k_{\perp}$ dependence weaker, see also related discussion in Ref.~\cite{Ducloue:2017kkq}. Finally, in the dilute (high-$k_{\perp}$) region, the nuclear effects vanish and $R_{p\mathrm{Pb}}\to 1$. This is in contrast to the leading order calculations where the leading order BK evolution changes the asymptotic shape of the dipole amplitude, and consequently one obtains $R_{p\mathrm{Pb}}\to 1$ at moderate transverse momenta $k_\perp \sim 5\dots 10$ GeV only at the initial condition~\cite{Lappi:2013zma,Albacete:2003iq}. At the LHC kinematics, a significantly large transverse momentum is needed in order to be sensitive to dipoles close to the initial condition where $X_g\approx X_0$, see Eq.~\eqref{eq:xg}, and as such $R_{p\mathrm{Pb}}$ would approach unity much more slowly in a complete LO calculation.

\section{Hadron level results}\label{sect:hadron_lvl_res}

In this section, we calculate the complete forward $\pi^0$ production cross sections for minimum bias proton-lead collisions at $\sqrt{s} = 8.16$ TeV. When calculating the impact parameter integral (recall that at each $b_{\perp}$ we use independently evolved dipole-nucleus amplitudes), we encounter regions where the saturation scale of the nucleus would be smaller than that of the proton. To avoid unphysically rapid growth of the nuclear size in this region we calculate the yield in $pA$ collisions at large $b_\perp$ following Ref.~\cite{Lappi:2013zma} as
\begin{equation}
    \dd N^{pA}(\bt) = N_\mathrm{bin}(\bt) \, \dd N^{pp} ,
\end{equation}
This corresponds to $R_{p\mathrm{Pb}}=1$ at large impact parameters.

\begin{figure*}[tb]

\subfloat[KCBK, parent-dipole coupling \label{subfig:KCBKfit1_spectra}]{%
  \includegraphics[width=0.48\textwidth]{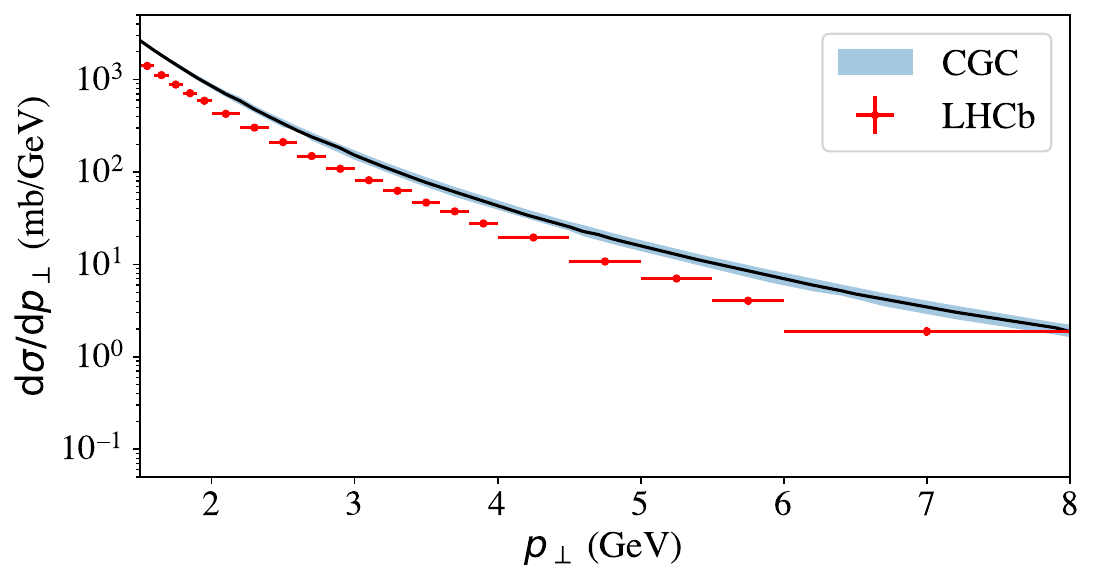}%
}\hfill
\subfloat[KCBK, Balitsky and smallest-dipole coupling \label{subfig:KCBKfit5_spectra}]{%
  \includegraphics[width=0.48\textwidth]{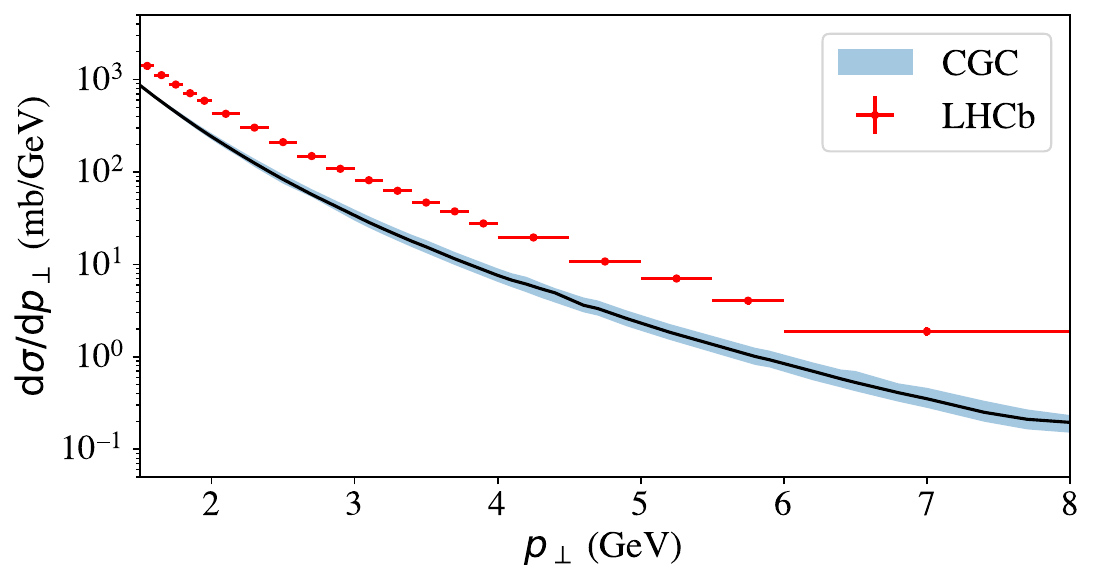}% 
  %\label{fig:kcbkbalitskyspectra}
}

\subfloat[ResumBK, parent-dipole coupling \label{subfig:ResumBKfit1_spectra}]{%
  \includegraphics[width=0.48\textwidth]{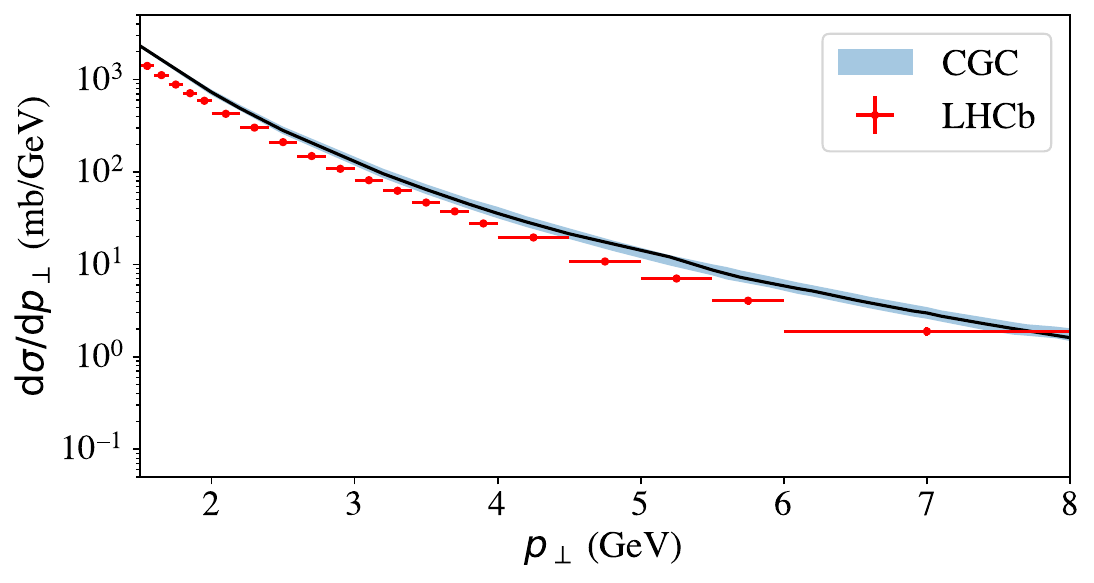}%
}\hfill
\subfloat[TBK, parent-dipole coupling \label{subfig:TBKfit1_spectra}]{%
  \includegraphics[width=0.48\textwidth]{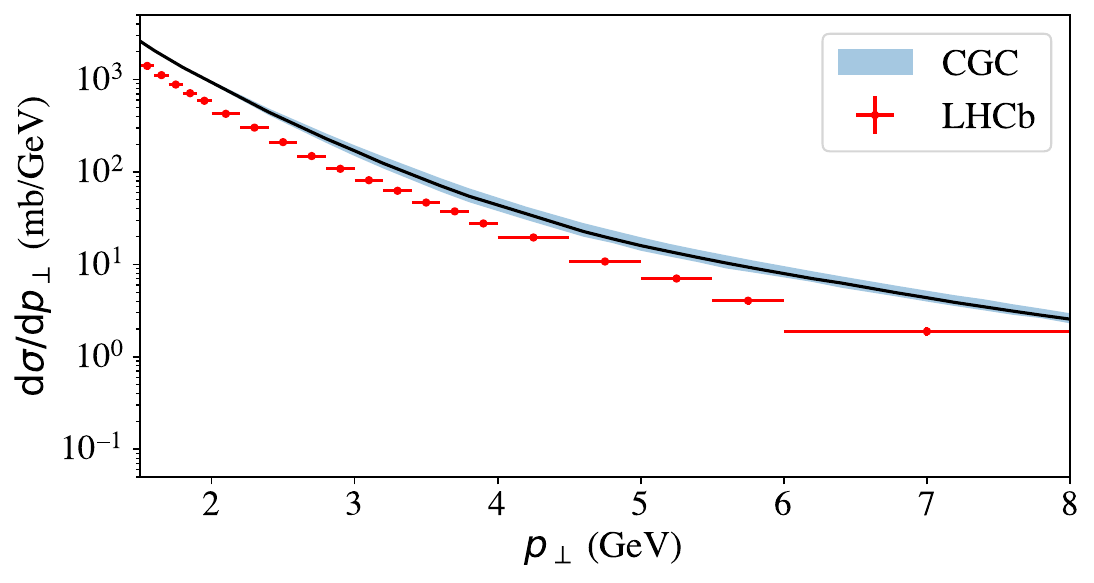}%
}

\caption{
Inclusive $p+\mathrm{Pb}\to \pi^0+X$ cross section as a function of pion transverse momentum, $p_\perp$, using different resummation schemes and different running coupling prescriptions in the BK evolution compared with the LHCb data from Ref.~\cite{LHCb:2022tjh}. Here, $\sqrt{s}=8.16$ TeV and the pion rapidity is $y=3$. The blue band is obtained by varying the factorization scale from $\mu=2p_\perp$ to $\mu=8 p_\perp$, and the central lines is obtained with $\mu=4 p_\perp$. The statistical and systematic uncertainties for the LHCb data are added in quadrature.
\label{fig:spectra_main}
}
\end{figure*}

The inclusive $\pi^0$ production cross sections at NLO as functions of pion transverse momentum compared to the LHCb data~\cite{LHCb:2022tjh} are shown in Figs.~\ref{fig:spectra_main} separately for the different resummation schemes used to approximate the full NLO BK evolution (KCBK, ResumBK and TBK). To demonstrate the sensitivity to the running coupling prescription employed in the BK evolution and to the related effects of the parameters in the initial condition, we also show the KCBK results using two different running coupling schemes employed in the NLO DIS fits in Ref.~\cite{Beuf:2020dxl}. Note that in the NLO impact factor, Eqs.~\eqref{NLOqq} and \eqref{NLOexpr}, the strong coupling is always evaluated at the scale given by the transverse momentum of the produced parton. The dependence on the factorization scale is illustrated by varying $\mu$  in the range $\mu=2p_{\perp},\ldots, 8p_{\perp}$.

The LHCb data shown for comparison span the rapidity range of $2.5<y<3.5$. We calculate the cross section at the central value,  $y=3$. Generically, the $p_{\perp}$ dependence of data is well captured in all the employed setups, except when the ``Balitsky+smallest dipole'' running coupling scheme (as defined in Ref.~\cite{Beuf:2020dxl}) is used in which case the spectrum falls more steeply than the data. Similarly, the overall normalization of the LHCb data is typically slightly overestimated, except in the Balitsky+smallest dipole running coupling setup in which case the normalization is also underestimated by a factor $\sim 2\dots 3$. As we show in Appendix.~\ref{sect:app_rc}, the overall normalization depends slightly (up to $\sim 50\%$) on the  running coupling prescription applied in the impact factor, but the shape of the $p_\perp$ spectra and the nuclear modification factor are insensitive to this scheme choice. We recall that in Ref.~\cite{Beuf:2020dxl} all resummation schemes for the BK equation applied here result in identical $\gamma^*p$ cross sections. This demonstrates that single inclusive hadron production at the LHC provides additional complementary constraints to the extraction of the non-perturbative initial condition for the dipole-proton scattering amplitude. Similar conclusions have  been made previously in studies involving exclusive vector meson production~\cite{Mantysaari:2022kdm,Mantysaari:2022bsp} and heavy quark production~\cite{Hanninen:2022gje} in DIS. This complementarity is due to the fact that the $\pi^0$ cross section in forward $pA$ collisions is sensitive to the dipole amplitude at different length scales than the total DIS cross section. In particular, small dipoles do not contribute to the structure function $F_2$, but on the other hand the shape of the dipole amplitude at small dipole size can have a dramatic effect, even rendering the (parton-level) cross section negative if the dipole amplitude vanishes faster than $r_\perp^2$ at small $r_\perp$ as discussed in Sec.~\ref{sect:dip} and in Ref.~\cite{Giraud:2016lgg}.  

We note that in Ref.~\cite{Beuf:2020dxl} more fits than the four that we use in Figs.~\ref{fig:spectra_main} are reported. We will further demonstrate the sensitivity of our results to the details of the NLO DIS fit later in this Section, but note that only the fits with parent dipole prescription used in Figs.~\ref{fig:spectra_main} have anomalous dimensions, $\gamma\approx 1$. This is preferred as the parton level cross section is then positive definite and the optical Glauber model extension to the dipole-nucleus scattering is more natural. For comparison, we show in Fig.~\ref{subfig:KCBKfit5_spectra} the result obtained using the KCBK evolution with the Balitsky+smallest dipole running coupling prescription, a fit that has the anomalous dimension of $\gamma=1.21$ at the initial condition. As discussed above, in the NLO evolution the anomalous dimension $\gamma$ does not change significantly~\cite{Lappi:2016fmu,Beuf:2020dxl}, and as such there remains a sensitivity to the initial anomalous dimension even at the LHC energies. Larger anomalous dimensions typically result in more steeply falling spectra as observed e.g. in Ref.~\cite{Lappi:2013zma} in a leading-order calculation. This is also evident in our NLO results when comparing Figs.~\ref{subfig:KCBKfit1_spectra} and \ref{subfig:KCBKfit5_spectra}: the spectra in Fig.~\ref{subfig:KCBKfit5_spectra} with $\gamma=1.21$ at the initial condition fall more steeply with $p_\perp$ than the LHCb data, compared to the case shown in Fig.~\ref{subfig:KCBKfit1_spectra} whose spectra describe the $p_\perp$ dependence of the LHCb measurement well with the initial anomalous dimension of $\gamma\approx 1$. Other fits with even larger $\gamma$ would yield $p_{\perp}$ spectra that fall even more steeply than the one shown in Fig.~\ref{subfig:KCBKfit5_spectra}, and hence these fits are not favored by the LHCb data. We note that, as shown in Ref.~\cite{Lappi:2013zma}, the leading order calculations require a larger $\gamma\sim 1.2$ in order to reproduce the $p_\perp$ dependence seen in the midrapidity proton-(anti)proton measurements at the LHC~\cite{CMS:2010tjh,ALICE:2012wos} and at Tevatron~\cite{CDF:2009cxa}.

\begin{figure*}[tb]

\subfloat[KCBK, parent-dipole coupling \label{subfig:KCBKfit1_RpA}]{%
  \includegraphics[width=0.48\textwidth]{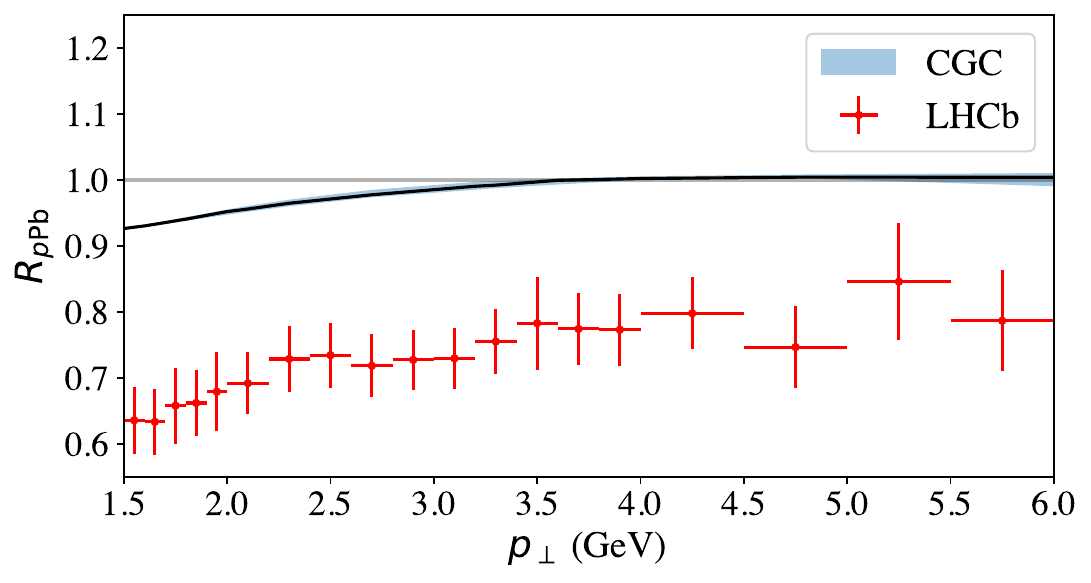}%
}\hfill
\subfloat[KCBK, Balitsky and smallest-dipole coupling \label{subfig:KCBKfit5_RpA}]{%
  \includegraphics[width=0.48\textwidth]{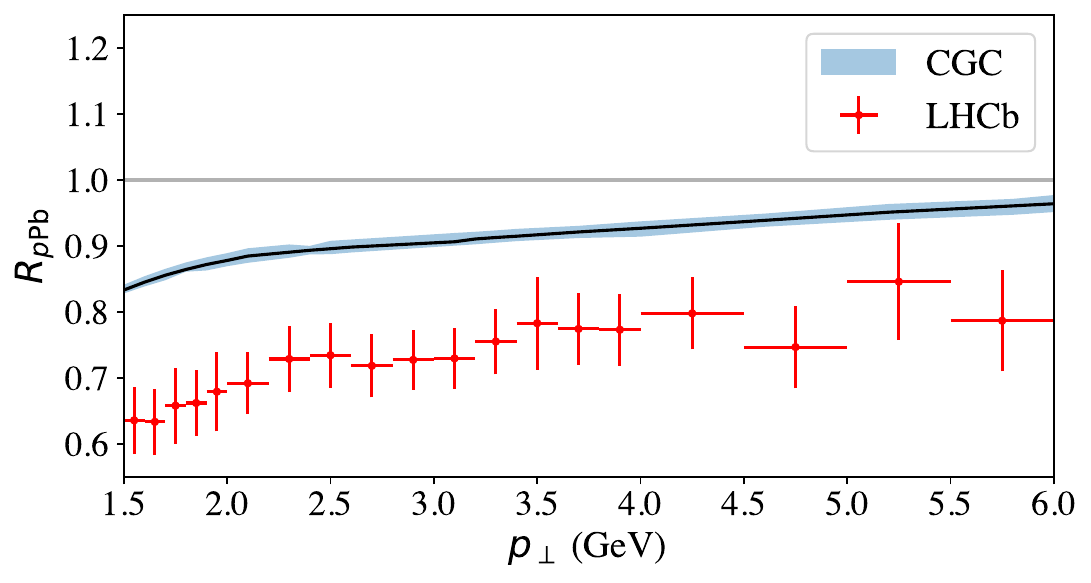}%
}

\subfloat[ResumBK, parent-dipole coupling \label{subfig:ResumBKfit1_RpA}]{%
  \includegraphics[width=0.48\textwidth]{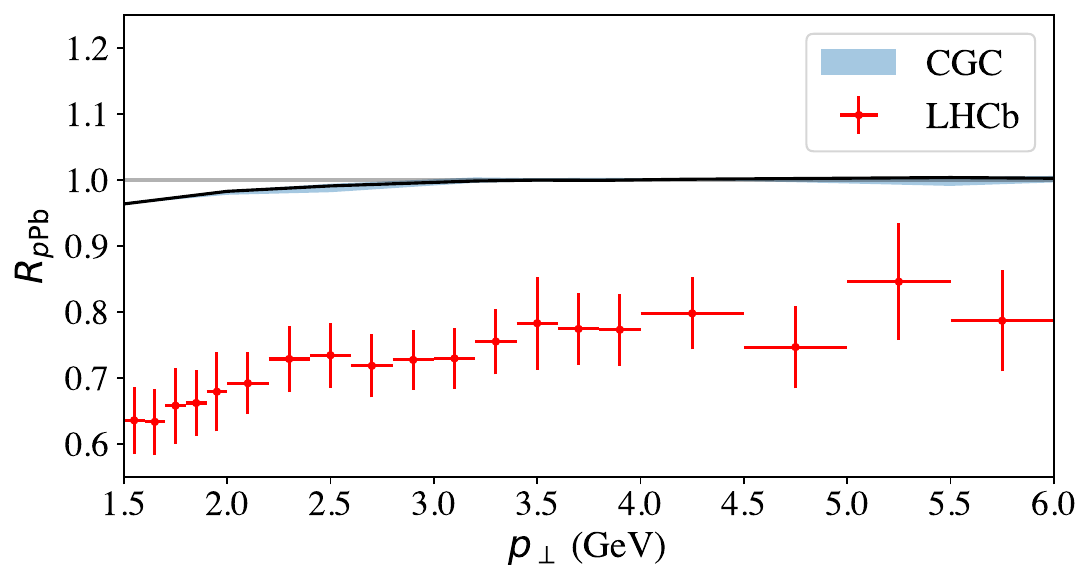}%
}\hfill
\subfloat[TBK, parent-dipole coupling \label{subfig:TBKfit1_RpA}]{%
  \includegraphics[width=0.48\textwidth]{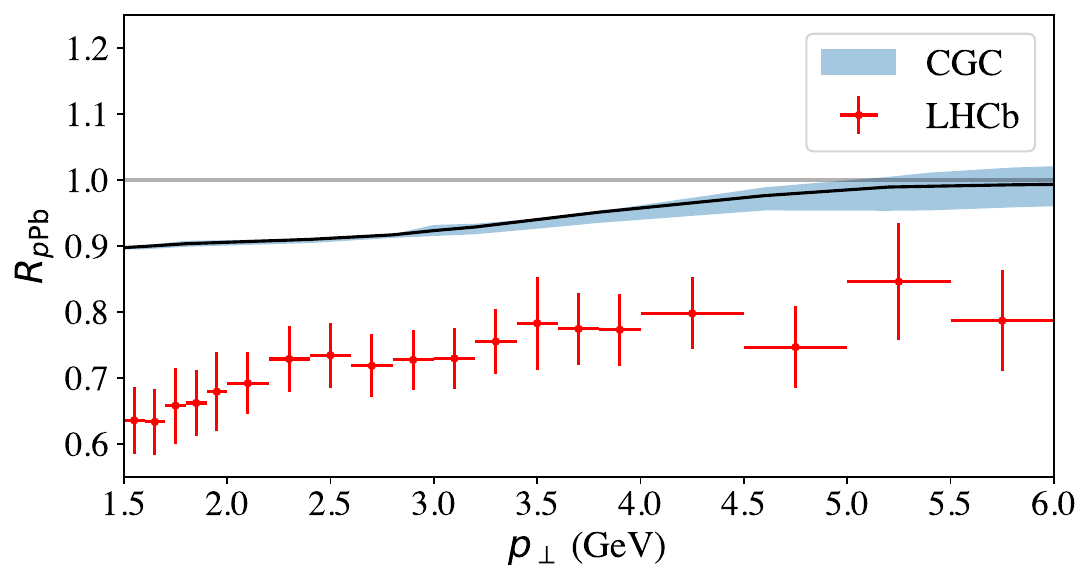}%
}

\caption{The nuclear modification factor $R_{p\text{Pb}}$ obtained using different resummation schemes and running coupling prescriptions in the BK evolution compared to the LHCb data~\cite{LHCb:2022tjh}. The results are calculated at $\sqrt{s}=8.16$ TeV and $y=3$. In each plot, the blue band with the black center line represents our numerical computation with $\mu \in \{2p_\perp,4p_\perp,8p_\perp\}$. The statistical and systematic uncertainties for the LHCb data are added in quadrature.}\label{fig:RpA_main}

\end{figure*}

Let us next consider the nuclear modification factor at the hadron level, defined as
\begin{equation}\label{RpA_hadron}
    R_{p\mathrm{A}} = \frac{\dd \sigma^{pA\to h+X}}{A\,\dd \sigma^{pp\to h+X}}\,.
\end{equation}
This was studied in the case of most central collisions ($b_{\perp}=0$) at the parton level in Sec.~\ref{sec:partonlevel}. The complete hadron-level results are shown in Figs.~\ref{fig:RpA_main} together with the corresponding LHCb measurements~\cite{LHCb:2022tjh}. The obtained nuclear modification factor is similar with the four NLO DIS fits applied. We find weak nuclear suppression at low hadron $p_\perp$ and obtain $R_{p\mathrm{Pb}}\to 1$ at moderate $p_\perp$, unlike in leading order calculations~\cite{Lappi:2013zma}.

The amount of nuclear suppression is significantly underestimated in all applied setups. The origin of this  can be traced back to the fact that the applied NLO fits for the dipole amplitude with $\gamma\sim 1$ (required to get positive definite parton level cross section) prefer relatively small proton transverse area $\sigma_0/2 = 10$ mb. This is significantly smaller than $\sigma_0/2 = 17$ mb obtained from the leading-order fits~\cite{Lappi:2013zma,Albacete:2010sy}. Since the saturation scale of the nucleus scales as $Q_{s,0,A}^2\sim \frac{\sigma_0}{2} A T_A(\bt)  Q_{s,0}^2$ according to the applied optical Glauber model (with $\gamma=1$), c.f. Eq.~\eqref{ICdip_pA}, we have a smaller difference between the proton and nuclear saturation scales compared to the leading-order case. This results in a weaker nuclear suppression at NLO. Furthermore, note that the applied fits with smaller $\sigma_0$ can be seen to be preferred by measurements of several other observables. For example, the momentum transfer dependence of the exclusive $\mathrm{J}/\psi$ production measurement from HERA~\cite{Alexa:2013xxa,ZEUS:2002wfj} corresponds to $\sigma_0/2 = 10$ mb~\cite{Caldwell:2009ke}, although the exact value depends on the assumed proton shape that cannot be exactly determined from the currently available data. Along the same line, the diffractive structure function data have been shown in Ref.~\cite{Lappi:2023frf} to require a very steep proton density profile when $\sigma_0/2 = 17$ mb is used in leading-order calculations, suggesting that somewhat smaller proton size could be preferred.

\begin{figure*}[htb]
\subfloat[Transverse momentum spectra]{ \label{fig:kcbkfit3_spectra}
  \includegraphics[width=0.48\textwidth]{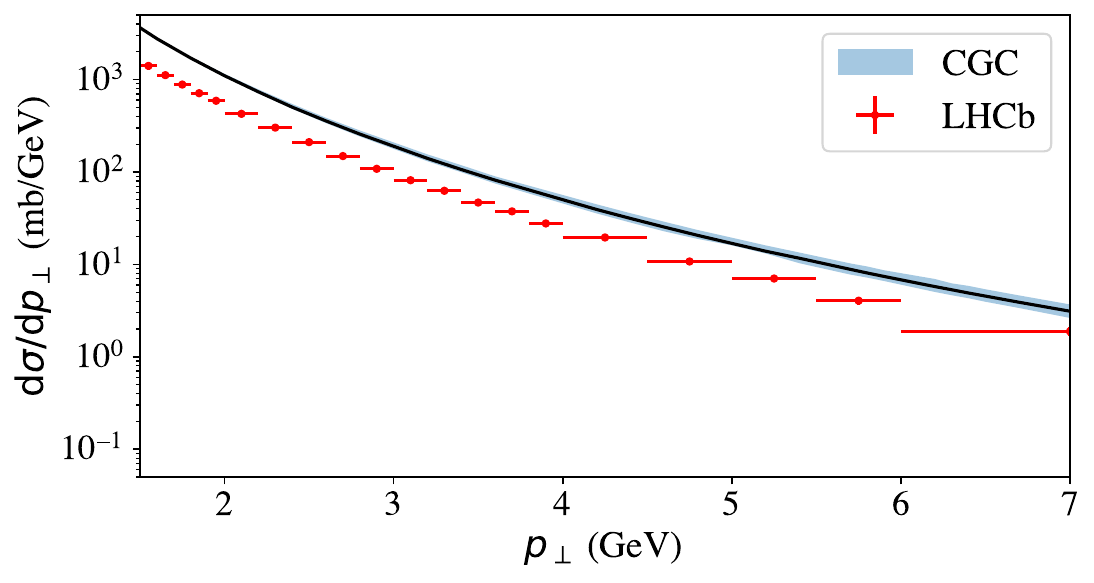}%
}\hfill
\subfloat[Nuclear modification factor]{%
\label{fig:kcbkfit3_rpa}
  \includegraphics[width=0.48\textwidth]{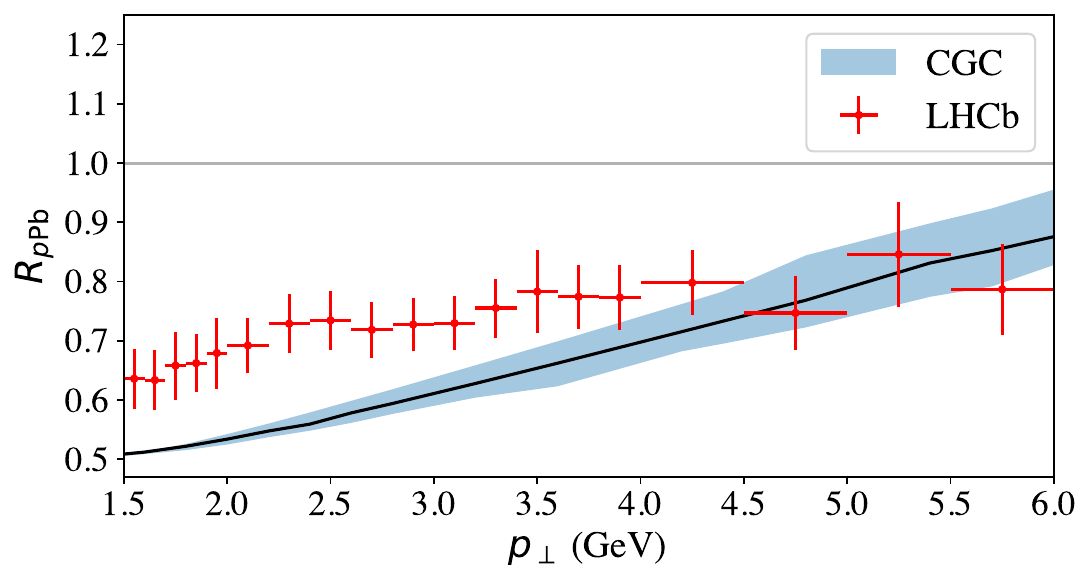}%
}
\caption{Inclusive $\pi^0$ spectra and nuclear modification factor computed using the KCBK resummation scheme for the BK evolution and a fit where the BK evolution starts at $X_0=1$. This particular fit has a larger proton transverse area, $\sigma_0/2=18.39$ mb. The results are compared with the LHCb data~\cite{LHCb:2022tjh}. The scale variation band is similarly generated with $\mu \in \{2p_\perp,4p_\perp,8p_\perp\}$.}
\label{fig:largersigma02}
\end{figure*}

In order to further demonstrate the sensitivity to the initial condition used in the small-$x$ evolution, we calculate the inclusive $\pi^0$ production cross section and the nuclear suppression factor using an initial condition parametrization that has a larger proton transverse area, $\sigma_0/2=18.39$ mb. This is the KCBK initial condition with parent dipole running coupling prescription where the BK evolution starts at $X_0=1$ in Ref.~\cite{Beuf:2020dxl}. The resulting pion spectrum and nuclear modification factor are shown in Figs.~\ref{fig:largersigma02}. Despite the relatively small difference one might expect from varying the starting point of the high-energy evolution, Figs.~\ref{fig:largersigma02} display clear differences when compared to the respective counterparts in Figs.~\ref{fig:spectra_main} and \ref{fig:RpA_main}. In large part, this is due to the differences in the BK initial condition's parameters that resulted from the fit to the HERA data in Ref.~\cite{Beuf:2020dxl}. In the large-$\sigma_0$ case of Figs.~\ref{fig:largersigma02}, we find that the overall cross section is typically overestimated by a factor $\sim 2$ as seen in the transverse momentum spectra shown in Fig.~\ref{fig:kcbkfit3_spectra}. This initial condition has an anomalous dimension $\gamma=1.21$, but unlike above when applied to fits where the BK evolution starts at $X_0=0.01$, the $p_\perp$ slope is only slightly too steep compared to the LHCb data. The fact that a larger $\gamma$ is not clearly disfavored in this case can be traced back to the fact that there is now an additional BK evolution from $X=1$ to $X=0.01$, in contrast to the previous case where the BK evolution only starts at $x=0.01$. Although the shape of the dipole amplitude does not change at asymptotically small $r_{\perp}$, the evolution can have a moderate effect on the shape of the dipole at intermediate dipole sizes~\cite{Beuf:2020dxl}, which can result in slight modifications on the $p_\perp$ slope.

The nuclear suppression factor shown in Fig.~\ref{fig:kcbkfit3_rpa} shows a much stronger suppression than those in Figs.~\ref{fig:RpA_main}, where fits with significantly smaller proton transverse areas, %$\sigma_0/2\sim 9$ mb
$\sigma_0/2 = 10$ mb, were used. As already discussed above, with larger $\sigma_0/2$ the relative difference between the proton and nuclear saturation scales is larger, resulting in stronger nuclear suppression. With this particular fit having $\sigma_0/2=18.39$ mb, we obtain an even stronger nuclear suppression at low pion transverse momentum than the level observed in the LHCb data. This demonstrates that $R_{p\mathrm{Pb}}$ measurements are very sensitive to the proton size parameter $\sigma_0/2$ (at least in the applied Optical Glauber model for the nucleus). This is in contrast to the DIS structure function case that are fitted to extract the initial condition for the BK evolution: in the linear regime the structure functions depend only on the product $\frac{\sigma_0}{2} Q_{s,0}^2$ and as a result significant correlations between these parameters can be expected. However, as no uncertainty estimate for the NLO DIS fits is currently available, it is not possible to quantify how the uncertainties in the fit parameters actually propagate to the nuclear modification factor calculated here. 
 
Finally, we study how the cross section and the nuclear modification factor depend on the pion rapidity which controls the amount of BK evolution in the calculation. These results can be seen as predictions for the future measurements. The LHCb collaboration has previously measured inclusive charged hadron production up to $y=4.3$~\cite{LHCb:2021vww}. Furthermore, in the future, the ALICE collaboration with the Forward Calorimeter (FoCal) detector upgrade should be able to perform measurements almost up to $y= 6$~\cite{ALICE:2023rol}.  

\begin{figure}
    \centering
    \includegraphics[width = \columnwidth]{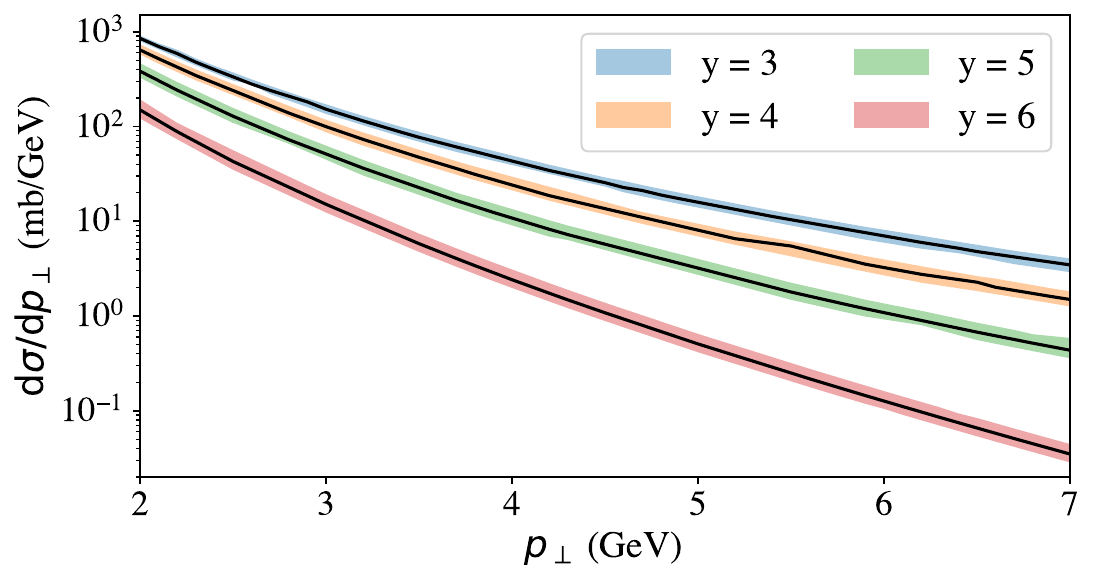}
    \caption{Inclusive $\pi^0$ spectra at rapidities $y=3,4,5,6$ in p+A collisions and $\sqrt{s}=8.16$ TeV. The calculation is performed with KCBK evolution and parent-dipole coupling. 
    The central values are obtained using a factorization scale $\mu=4p_\perp$ and the bands correspond to variation $\mu=2p_\perp \dots 8p_\perp$.}
    \label{fig:KCBK_ys_spectra}
\end{figure}

\begin{figure}
    \centering
    \includegraphics[width = \columnwidth]{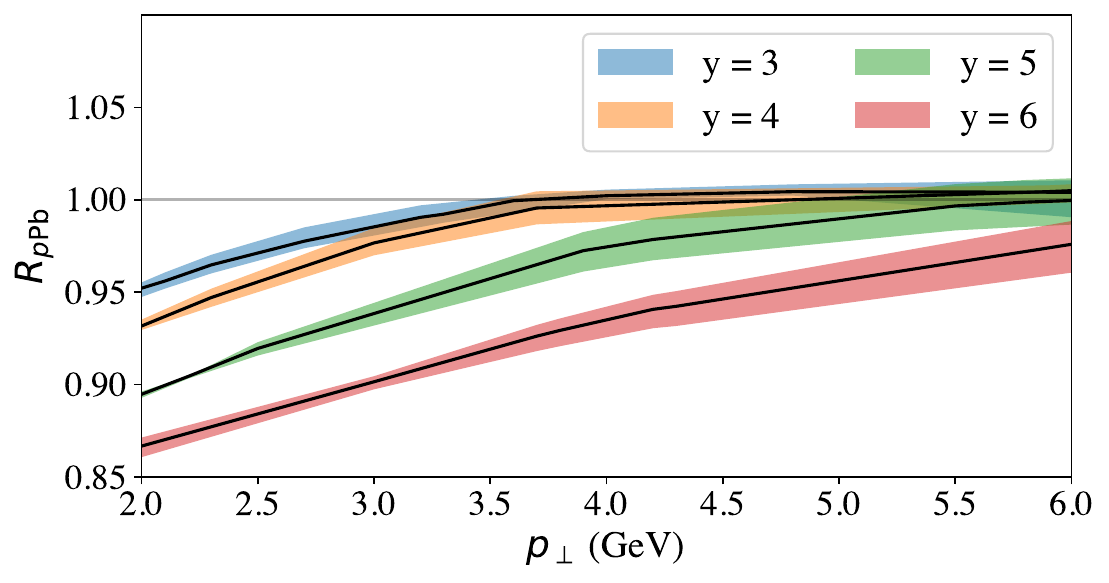}
    \caption{Nuclear suppression factor $R_{p\text{Pb}}$ for $\pi^0$ production at rapidities $y=3,4,5,6$ and $\sqrt{s}=8.16$ TeV. The calculation is performed with KCBK evolution and parent-dipole coupling. The central values are obtained using a factorization scale $\mu=4p_\perp$ and the bands correspond to variation $\mu=2p_\perp \dots 8p_\perp$.
    }
    \label{fig:KCBK_ys_RpA_NLO}
\end{figure}

The inclusive $\pi^0$ spectra at rapidities $y=3,4,5,6$ are shown in Fig.~\ref{fig:KCBK_ys_spectra}, and similarly the nuclear modification factors at the same values of rapidity are shown in Fig.~\ref{fig:KCBK_ys_RpA_NLO}. Here, we only use the KCBK resummation scheme for the BK evolution with the parent dipole running coupling prescription in the case where the evolution starts at $X_0=0.01$, i.e. with the proton size $\sigma_0/2=9.74$ mb and $\gamma=0.98$, that was used also in Figs.~\ref{subfig:KCBKfit1_spectra} and~\ref{subfig:KCBKfit1_RpA}. A reason for this particular choice of resummation scheme, running coupling prescription and evolution starting point lies in the fact that this combination results in $\gamma<1$ according to the fit in Ref.~\cite{Beuf:2020dxl}. This implies that the cross section is positive definite as long as the unsubtracted scheme is employed in conjunction with the momentum-space running coupling prescription for the hard factor \cite{Ducloue:2017dit}. As discussed above, the magnitude of the nuclear suppression obtained depends strongly on the proton transverse area on which there is significant variation between the available DIS fits, but the rapidity dependence should not depend strongly on the proton size. The $\pi^0$ production cross section is heavily suppressed at large rapidities as the quark and gluon PDFs vanish rapidly when approaching the kinematical boundary $x_p=1$. The nuclear suppression is found to be stronger for more forward rapidities due to the larger saturation scale of the nucleus probed in such kinematics. In the moderately large transverse momentum region, $p_\perp\gtrsim 5$ GeV, we again get $R_{p\mathrm{Pb}}\to 1$. This feature is independent of the pion rapidity and qualitatively different from what is seen in leading-order calculations~\cite{Lappi:2013zma}. These observations qualitatively agree with the trends found in the LHCb measurements of charged hadron production at different rapidities (with $\sqrt{s}=5$ TeV)~\cite{LHCb:2021vww}, although the rapidity dependence obtained in this work is weaker. 

\section{Conclusion}\label{sect:conclusion}

In this work, we have performed the first calculation of single inclusive $\pi^0$ production completely at next-to-leading accuracy in the Color Glass Condensate approach consistently with the DIS structure function data. Forward particle production processes in p+A collisions probe nuclear wave functions in the small-$x$ regime. As a result, significant saturation effects can be expected in this kinematics. The developments presented in this work enable precision studies of gluon saturation at the smallest values of $x$ accessible with the currently available collider energies.

First, we extended the results of Ref.~\cite{Ducloue:2017dit} and calculated contributions from all partonic channels to the inclusive particle production cross section in the unsubtracted scheme. This allowed us to consistently use the dipole-target scattering amplitudes extracted from the HERA structure function data at NLO in Ref.~\cite{Beuf:2020dxl}. These results are shown in Eqs.~\eqref{NLOexpr}. Then, we demonstrated that the NLO corrections both to the small-$x$ evolution and to the impact factor have qualitatively significant effects on the nuclear modification factor in inclusive particle production processes. A distinct qualitative feature of the fully consistent NLO results presented in this work is that, at all rapidities, the $R_{p\mathrm{Pb}}$ approaches $1$ already at moderate hadron transverse momentum, $p_{\perp}=5\dots 10$ GeV. This is in contrast to the LO results for which the $R_{p\mathrm{Pb}}$ remains suppressed up to very large $p_{\perp}$ in the LHC kinematics~\cite{Lappi:2013zma}.

When both the impact factor and the small-$x$ evolution are consistently taken at NLO accuracy, we obtain a nuclear modification factor and inclusive $\pi^0$ spectra that qualitatively agree with the LHCb data as shown in Figs.~\ref{fig:spectra_main},~\ref{fig:RpA_main} and~\ref{fig:largersigma02}. Depending on the details of the applied DIS fit parametrizing the initial condition for the high-energy evolution, in particular on the extracted proton (gluonic) transverse area, our results for the nuclear modification factor $R_{p\mathrm{Pb}}$ either under- or overestimate the observed nuclear suppression in the inclusive forward $\pi^0$ production. Overall, the shape of the transverse momentum spectra is found to further constrain the DIS fits, disfavoring parametrizations with large anomalous dimensions $\gamma$ corresponding to steeply falling dipole-proton amplitudes at small dipole sizes. We note that similar additional constraints to the non-perturbative initial condition have recently been obtained when other DIS data sets have been included in the analysis in addition to the total cross section~\cite{Hanninen:2022gje,Mantysaari:2022kdm,Mantysaari:2022bsp}. 

Because of the strong sensitivity to the initial condition of the BK evolution it is necessary to perform global analyses in the CGC framework that include a variety of different DIS and p+A observables with rigorously propagated uncertainties, in order to obtain robust signals of non-linear QCD dynamics at currently available collider energies. The feasibility of such analyses simultaneously including both the total DIS cross section and forward LHC data has been demonstrated in this work where the first consistent NLO calculation has been presented, and the importance of a consistent treatment of different NLO corrections has been illustrated.

%At this point, we  and the LHCb measurements \cite{LHCb:2022tjh} to be a result of the parameters in the initial condition for the small-$x$ evolution, which were fitted to the HERA structure function data only \cite{Beuf:2020dxl}. The results in this work highlight the need to perform global analyses in the CGC framework that include a variety of different DIS and p+A observables with rigorously propagated uncertainties, in order to obtain robust signals of non-linear QCD dynamics at currently available collider energies. 

\begin{acknowledgements}

We thank B. Ducloué for providing the code developed in Ref.~\cite{Ducloue:2017dit}. We thank T. Lappi and J. Jalilian-Marian for
discussions and T. Boettcher for clarifying the normalization in LHCb measurements. This work was supported by
the Research Council of Finland, the Centre of Excellence
in Quark Matter and Projects No. 338263 and No. 346567.
This work was also supported under the European Union’s
Horizon 2020 research and innovation programme by
the European Research Council (ERC, Grant Agreement
No. ERC-2018-ADG-835105 YoctoLHC) and by the
STRONG-2020 project (Grant Agreement No. 824093).
Computing resources from the Finnish Grid and Cloud
Infrastructure (persistent identifier \texttt{urn:nbn:fi:research-infras-2016072533}) were used in this
work. The content of this article does not reflect the official
opinion of the European Union and responsibility for the
information and views expressed therein lies entirely with
the authors.

\end{acknowledgements}

\bibliographystyle{JHEP-2modlong}
\providecommand{\href}[2]{#2}\begingroup\raggedright\endgroup

\appendix
\section{Effects of Running Coupling Prescriptions in the Impact Factor}\label{sect:app_rc}

In this Appendix, we compare the results obtained using different running coupling prescriptions in the NLO impact factor, Eqs.~\eqref{NLOqq} and~\eqref{NLOexpr}, while using the same parent-dipole prescription in the small-$x$ evolution. For simplicity we only consider $\pi^0$ production in the most central $b_{\perp}=0$ case. Specifically, the running coupling prescriptions considered here are:
\begin{enumerate}
    \item[(i)] \emph{Fixed coupling,} with $\bar{\alpha}_s = \alpha_sN_c/\pi = 0.2$.
    \item[(ii)] \emph{Parent-dipole prescription,} given by the transverse separation between the positions of the incoming partons (before the primary parton emission) in the amplitude and the complex-conjugate amplitude. The coupling constant in term of a transverse separation, $\rt$, is given by \cite{Beuf:2020dxl}
    \begin{align}\label{aspos}
        &\alpha_s(\rt) = \frac{4\pi}{\beta_0\,\ln\left[\left(\frac{\mu_0^2}{\lqcd^2}\right)^{1/c} + \left(\frac{4C^2}{\rt^2\lqcd^2}\right)^{1/c} \right]^c} \, ,
    \end{align}
    where $\beta_0 = (11N_c-2N_f)/3$ and $\lqcd = 0.241$ GeV. Here, recall that we include only the three lightest quark flavors: $N_f=3$. For the parameters characterizing the infrared behavior of the coupling constant, we follow the convention from \cite{Beuf:2020dxl} and choose $\mu_0/\lqcd = 2.5$ and $c=0.2$.  
    \item[(iii)] \emph{Momentum-space prescription,} given by the transverse momentum, $\kt$, of the produced parton, such that
    \begin{align}\label{asmom}
        &\alpha_s(\kt) = \frac{4\pi}{\beta_0\,\ln(k_{\perp}^2/\lqcd^2)}\,,
    \end{align}
    with the same values for $\beta_0$ and $\lqcd$.
    \item[(iv)] \emph{Mixed-space prescription,} which is identical to the momentum-space prescription, except for the $qq$-channel terms proportional to $J$ and $J_v$, c.f. Eqs.~\eqref{NLOqq}, \eqref{Jqq} and \eqref{Jvqq}. In those terms, the coupling constant runs with the separation, $x_{\perp}$, between the two partons after the primary parton emission, according to Eq.~\eqref{aspos}. This is the prescription employed in \cite{Ducloue:2017dit}. In this work, we find that a finite result can only be achieved when we restrict position-space prescriptions to these two specific $J$ and $J_v$ terms; in other terms, any linear combination of transverse positions involving the daughter dipole would contain an undesirable ``fake potential'' contribution, which will be discussed shortly. Hence, in most terms, we resort to the momentum-space prescription given in Eq.~\eqref{asmom}.
\end{enumerate}

\begin{figure}[tb]
    \centering
    \includegraphics[width = \columnwidth]{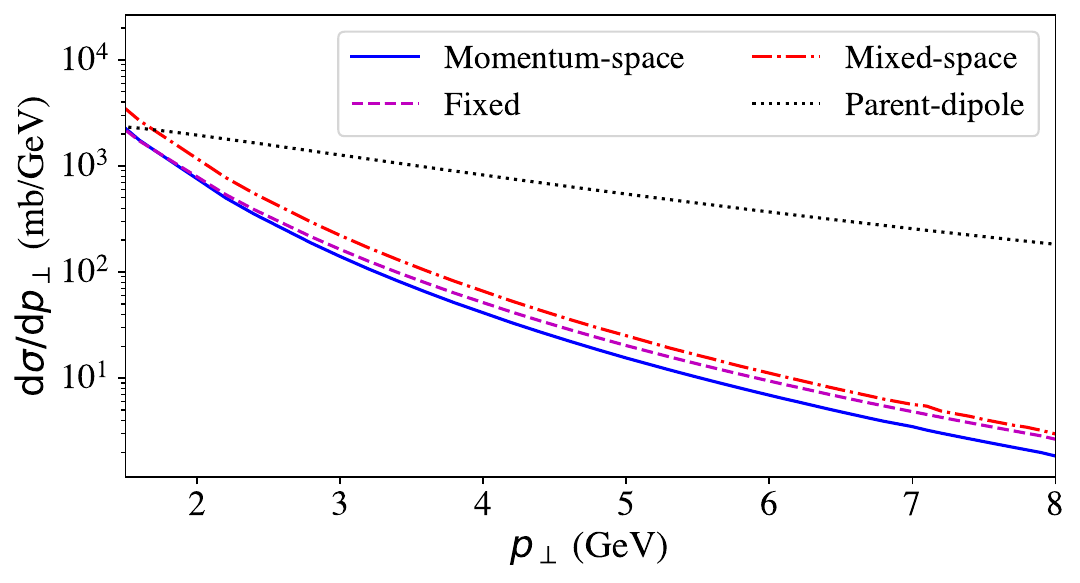}
    \caption{A plot of $p$Pb spectra computed using different running coupling prescriptions in the impact factor, while using the parent-dipole prescription in the BK evolution. }
    \label{fig:KCBK_rc_spectra}
\end{figure}

The $\pi^0$ spectra for $p$Pb collisions are plotted in Fig.~\ref{fig:KCBK_rc_spectra} using the four different running coupling prescriptions in the impact factor. There, we see that the momentum-space and mixed-space prescriptions result in spectra that differ from each other by an overall normalization factor, but the shape is almost identical. On the other hand, when the running coupling effects are neglected in the fixed coupling case, one obtains a somewhat harder spectrum as expected.  Overall differences in the cross sections calculated using these three schemes are within 50\%. In contrast, a stark difference is observed when the parent-dipole prescription is used. In that case, the spectrum is much greater in magnitude and displays an exponential decay with $p_{\perp}$ (note the logarithmic scale in Fig.~\ref{fig:KCBK_rc_spectra}). The effect is also clearly visible in the nuclear modification factor. 

\begin{figure}[b]
    \centering
    \includegraphics[width = \columnwidth]{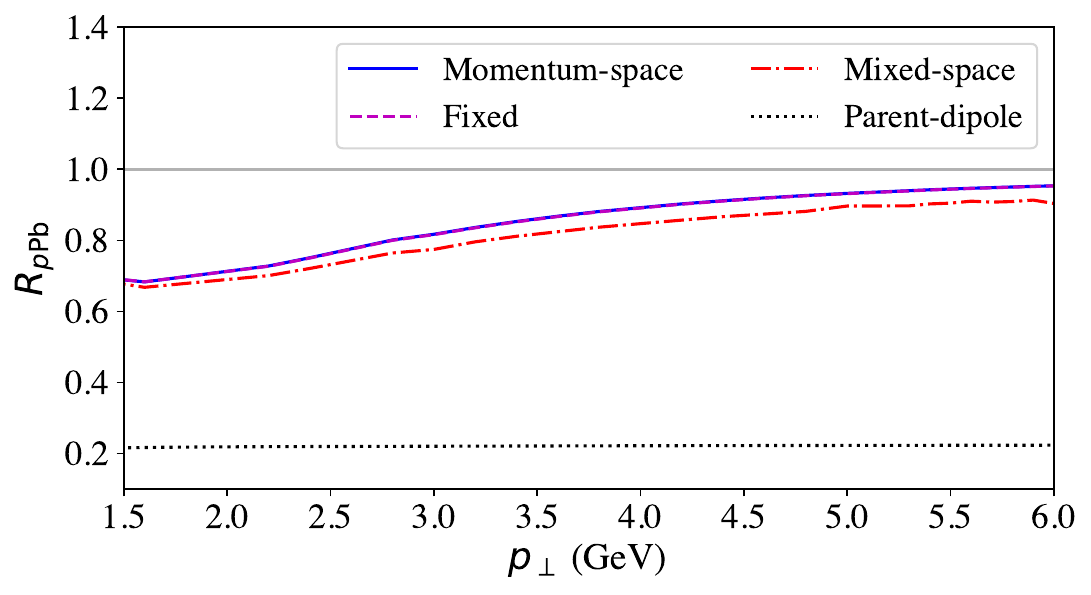}
    \caption{A plot of nuclear modification factor, $R_{p\text{Pb}}$, computed using different running coupling prescriptions in the impact factor, while using the parent-dipole prescription in the BK evolution. Note that the blue and magenta lines, corresponding respectively to the momentum-space and fixed coupling prescriptions, lie on top of each other. 
    }
    \label{fig:KCBK_rc_RpA}
\end{figure}

Fig.~\ref{fig:KCBK_rc_RpA} shows the $R_{p\text{Pb}}$ for different running coupling prescriptions in the impact factor. Here, we see that the results are almost identical for all the prescriptions except for the parent-dipole, whose nuclear modification factor simply remains a constant.

The qualitatively different behavior obtained with the parent-dipole scheme is due to the  ``fake potential'' problem discovered in Ref.~\cite{Ducloue:2017dit}. The main idea is that the dependence of the coupling constant on a transverse position changes the Fourier transform of the dipole amplitude, creating extra unphysical terms that dominate the cross section. The constant in the parent-dipole $R_{p\text{Pb}}$ is simply the ratio between the unphysical terms that would result from the $p$Pb and $pp$ initial conditions.

Finally, it is worth noting that employing any position-space prescription in any of the terms besides the $J$ and $J_v$ terms in the $qq$ channel discussed above would also produce a fake-potential contribution, resulting in the spectra and nuclear modification factor identical to those of the parent-dipole prescription. 

With the relatively minor differences among all the prescriptions that are free of the fake-potential contribution, we decide to employ the momentum-space prescription for the impact factor throughout the main sections of this work.

\end{document}